\documentclass[a4paper,12pt]{article}

\usepackage{amssymb,amsmath,bm}
\usepackage[totalwidth=17cm,totalheight=24cm]{geometry}
\usepackage{graphicx}

\usepackage{mathtools} 
\usepackage{IEEEtrantools}
\usepackage[usenames, dvipsnames]{color}
\usepackage{comment} 
\usepackage{fourier} 
\usepackage{accents} 
\usepackage{datetime} 

\usepackage{import}

\newcommand{\C}{{\mathbb C}}
\newcommand{\R}{{\mathbb R}}

\newcommand{\id}{{\mathbb 1}}
\newcommand{\im}{{\rm i\,}}

\usepackage{accents} 

\newcommand{\utilde}[1]{\underaccent{\widetilde}{#1}}

\newcommand{\ga}{\alpha} 						

\newcommand{\gat}{\widetilde{\alpha}}
\newcommand{\gb}{\beta}							

\newcommand{\gc}{\gamma}						

\newcommand{\gd}{\delta}						

\newcommand{\eps}{\epsilon}						

\newcommand{\epst}{\widetilde{\epsilon}}
\newcommand{\epsut}{\utilde{\epsilon}}

\newcommand{\tht}{\theta}						

\newcommand{\thtt}{\widetilde{\theta}}

\newcommand{\gl}{\lambda}						

\newcommand{\gs}{\sigma}						

\newcommand{\go}{\omega}						

\newcommand{\gO}{\Omega}

\newcommand{\ct}{\widetilde{c}}					

\newcommand{\J}{\text{J}}						
\newcommand{\Jt}{\widetilde{\text{J}}}

\newcommand{\Lc}{\mathcal{L}}

\newcommand{\Nt}{\widetilde{N}}					

\newcommand{\from}{\colon}

\newcommand{\pa}{\partial}
\newcommand{\W}{\wedge}

\newcommand{\be}{\begin{eqnarray}}
\newcommand{\ee}{\end{eqnarray}}

\begin{document}
 \pagestyle{plain}
\title{Topological field theories of 2- and 3-forms \\ in six dimensions}
\author{Yannick Herfray${}^{(1),(2)}$ and Kirill Krasnov${}^{(1)}$ \\ {}\\
{\small \it ${}^{(1)}$School of Mathematical Sciences, University of Nottingham, NG7 2RD,
UK}
\\
{\small \it ${}^{(2)}$ Laboratoire de Physique, ENS de Lyon,  46 all\'ee d'Italie, F-69364 Lyon Cedex 07, France}} 

\date{May 2017}
\maketitle

\begin{abstract}\noindent We consider several diffeomorphism invariant field theories of 2- and 3-forms in six dimensions. They all share the same kinetic term $BdC$, but differ in the potential term that is added. The theory $BdC$ with no potential term is topological --- it describes no propagating degrees of freedom. We show that the theory continues to remain topological when either the $BBB$  or $C\hat{C}$ potential term is added. The latter theory can be viewed as a background independent version of the 6-dimensional Hitchin theory, for its critical points are complex or para-complex 6-manifolds, but unlike in Hitchin's construction, one does not need to choose of a background cohomology class to define the theory. We also show that the dimensional reduction of the $C\hat{C}$ theory to three dimensions, when reducing on $S^3$, gives 3D gravity. 
\end{abstract}

\section{Theories of 2- and 3-forms in 6 dimensions}

We call a diffeomorphism invariant theory topological if it describes no propagating degrees of freedom. This means that the space of solutions modulo gauge (on a compact manifold) is at most finite dimensional. Topological field theories of Schwarz type \cite{Schwarz:1978cn}, \cite{Schwarz:1979ae} with Lagrangians of the type $C_p d C_{n-p-1}$ on an $n$-dimensional manifold $M^n$, where $C_p\in\Lambda^p(M^n)$ are $p$-forms, are very-well known. The partition function of such a theory is a variant of Ray-Singer analytic torsion of the manifold. 

Depending on the types of forms used and on the dimension of the manifold, it may be possible to add to the Lagrangian other potential-type terms without breaking the diffeomorphism invariant character of the theory. Depending on the specifics of such a modification, the theories obtained this way may or may not remain topological. In this article we study a set of examples of theories obtained this way, in the setting of a 6-dimensional manifold. A 7-dimensional example of this sort has recently been studied in \cite{Krasnov:2017uam}.

The topological theory that provides the kinetic term for all our constructions is that with the Lagrangian $BdC$ with $B\in\Lambda^2(M^6)$ and $C\in\Lambda^3(M^6)$. The main purpose of the paper is to point out that this obviously topological theory admits a set of modifications that keep its topological character unchanged. Our other purpose is to further characterise these topological theories by computing their dimensional reduction to 3-dimensions. For one of our topological theories in 6D the result of the dimensional reduction is 3D gravity. 

This study can be viewed as a continuation of the work \cite{Herfray:2016std}. That work considered the Hitchin functional \cite{Hitchin:2000jd} for 3-forms in 6 dimensions, in the setting where the 6-manifold was taken to be the total space of an ${\rm SU}(2)$ bundle over a 3-dimensional base. The 3-form considered was the Chern-Simons 3-form for a connection in this bundle. It was observed that the Hitchin functional reduces to the action for 3D gravity, in the so-called pure connection formalism. This construction can also be viewed as an explanation for why the pure connection formulation of 3D gravity exists. The starting point for this work was our desire to find a 6D functional whose dimensional reduction to 3D, when the internal manifold is taken to be the group ${\rm SU}(2)$, would give the first order formulation of 3D gravity with its frame field and the spin connection as the independent fields. This led us to consider the actions we study in the present paper. However, we also realised that apart from the action whose dimensional reduction gives 3D gravity, there are other natural and interesting actions to consider. Further, it seems that the process of "deforming" the topological theory of Schwarz type by adding to it "potential" terms, i.e. terms not involving any derivatives, has not been studied in the literature. There are some very interesting theories that can be obtained this way, as we aim to show in the present paper in the setting of 6 dimensions. 

We now consider some possible theories that can be obtained.

\subsection{Symplectic manifolds}

In the setting of 6 dimensions, the simplest possibility to consider is to add to the Lagrangian the term $B^3$, assuming that $B$ is non-degenerate so that this top form is non-zero. Thus, consider
\be\label{action-symp}
S_{symp}[B,C]=\int B\wedge dC +\frac{1}{3} B\wedge B\wedge B.
\ee 
The Euler-Lagrange equations that follow by extremising this action functional are
\begin{equation*}
dB=0, \qquad dC=- B\wedge B.
\end{equation*}

The 3-form field $C$ plays the role of the Lagrange multiplier imposing the condition that $B$ is closed. Critical points of this theory are therefore symplectic manifolds, with the additional constraint that $B\wedge B$ is exact. 

We note that the numerical coefficient that could have been put in front of the second term in (\ref{action-symp}) can be absorbed by the simultaneous redefinition of the $B$ and $C$ fields. 

In this paper we are mostly interested by the following question: does the theory starts to have any local degrees of freedom after deformation by a given `potential term'? As we shall see below, the `BBB deformed' theory (\ref{action-symp}) continues to remain topological, i.e. there are no propagating degrees of freedom. 

\subsection{Complex manifolds}

In the previous example we added to the Lagrangian the top form constructed as the cube of the 2-form $B$. It is natural to wonder if we can repeat the same construction with $C$, i.e adding a top form constructed from $C$. The most straightforward attempt does not however give anything as the wedge product $C\wedge C$ vanishes. 

Nevertheless, as was first described by Hitchin in \cite{Hitchin:2000jd}, there is another natural top form that can be constructed. As described in this reference, a generic (or stable in Hitchin's terminology) 3-form in 6 dimensions defines an endomorphism of the cotangent bundle that squares to plus or minus the identity. The most interesting case is that when we get the minus sign, i.e. an almost complex structure. Let us denote this almost complex structure by $J_C$. We can then apply $J_C$ to all the 3 slots of $C$ and obtain another 3-form, denoted by $\hat{C}$. More details on this construction are to be given below. We can then consider the following variational principle  
\be\label{action-complex}
S_{compl}[B,C] = \int B\wedge dC + \frac{1}{2} C\wedge \hat{C}.
\ee
Similar to (\ref{action-symp}), any parameter that may have been put in front of the second term can be absorbed by a field rescaling.

The Euler-Lagrange equations that describe extrema of this functional are
\begin{equation*}
dC=0,  \qquad dB = \hat{C}.
\end{equation*}
In particular the second equation implies that 
\begin{equation*}
d\hat{C}=0,
\end{equation*}
and thus the 3-form is closed and "co-closed" in the sense of the Hitchin story. As was shown by Hitchin \cite{Hitchin:2000jd}, this implies that the eigenspace distribution for $J_C$ is integrable. 

Our equations are stronger than those in the case of Hitchin because the second equation $dB = \hat{C}$ says that the 3-form $\hat{C}$ is exact, not just closed. In contrast, Hitchin starts with the functional which is just our second term in (\ref{action-complex}), but considers the variational principle in a fixed cohomology class of $C$, varying $C$ by an exact form. The resulting Euler-Lagrange equation in this case simply says that $\hat{C}$ is closed. 

What we have done is to impose the condition that $C$ is closed with a Lagrange multiplier, this is our first term in (\ref{action-complex}). Now variations over $C$ are not restricted to lie in a fixed cohomology class, and as the result we obtain a stronger condition on $\hat{C}$. The justification for our construction is that the theory (\ref{action-complex}) is "background independent" in the sense that one does not need to fix any background structure (like a cohomology class) to define it. 

As we shall see below, the theory (\ref{action-complex}) remains topological, i.e. there are no propagating DOF. This theory is particularly interesting because its dimensional reduction to 3 dimensions gives 3D gravity, see below. 

\subsection{Nearly-K\"ahler manifolds}

We now put together the two constructions above. Thus, we consider
\be\label{action-nK}
S_{n-kahler}[B,C] = \int B\wedge dC + \frac{1}{2} C\wedge \hat{C} + \frac{1}{3} B\wedge B\wedge B.
\ee
The numerical coefficients that could have been put in front of the second and third terms can be absorbed by the field $B,C$ redefinition, and multiplying the action by an overall constant. So, the only parameter in the above theory is the coefficient in front of the action, or, in physics terminology, the Planck constant. This parameter only matters in the quantum theory, where the partition function of the theory will depend on it. The Euler-Lagrange  equations that describe the extrema of the functional (\ref{action-nK}) are 
\begin{equation*}
    dC = - B\wedge B, \qquad dB = \hat{C},
\end{equation*}
and so the 3-form is no longer closed in this version of the theory. It is known, see Hitchin \cite{Hitchin:2001rw} Theorem 6, that a cone over such a structure gives a $G_2$-structure in 7 dimensions that defines a manifold of holonomy $G_2$. It then follows that such a structure corresponds to a nearly K\"ahler manifold. Note that normalisations of $B,C$ are different from the normalisations in \cite{Hitchin:2001rw}, with the equations naturally arising in the cone case being $dB=3\hat{C}, dC=-2B\wedge B$. The corresponding 3-form in 7-dimensions is $\Omega = t^2 dt \wedge B + t^3 \hat{C}$, and the dual form is ${}^*\Omega = t^3 dt \wedge C - (t^4/2) B\wedge B$. 

The most interesting point about the theory (\ref{action-nK}) is that the data $B,C$ in this case define a metric on $M^6$. This is of course compatible with the above statement that a cone over such a 6-manifold gives a 7-dimensional manifold of holonomy $G_2$. To see the appearance of the metric we note that taking the exterior derivative of the first field equation above we have $dB\wedge B=0$. Now, using the second equation we get
\begin{equation*}
    \hat{C} \wedge B=0.
\end{equation*}
As we will show below, this equation implies that the almost complex structure constructed from $\hat{C}$ (which is the same as $J_C$), is compatible with the almost symplectic structure $B$ in the sense that $B(\cdot, J\cdot)$ is a symmetric tensor. This metric, whose construction uses both $B$ and $C$ then gives $M^6$ the structure of a nearly-K\"ahler manifold. 

We remind the reader that nearly K\"ahler manifolds in dimension six have special properties. They are Einstein spaces of positive scalar curvature. They admit a spin structure and admit real Killing spinors. In fact the Einstein property follows directly from the fact that such manifolds admit Killing spinors. A useful reference on nearly K\"ahler manifolds (and much more) is \cite{Alexandrov:2004cp}, see in particular section 4.2 and references therein. 

So, the field equations for the theory (\ref{action-nK}) imply that the metric constructed from $B,C$ is Einstein, and in this sense (\ref{action-nK}) can be viewed as a gravity theory. This theory is thus a 6D analog of the 7D theory considered in \cite{Krasnov:2016wvc}, \cite{Krasnov:2017uam} in the sense that both of these theories are theories of differential forms whose equations imply that the metric that is constructible from the differential forms is Einstein. The theory (\ref{action-nK}) is however unlikely to be topological, see below. 

\subsection{Remarks}

In \cite{Hitchin:2004ut} Hitchin described a generalisation of the volume functional $C\wedge \hat{C}$ to all odd or even polyforms in 6D. There is thus a generalisation of all 3 theories (\ref{action-symp}), (\ref{action-complex}) and (\ref{action-nK}) to polyforms, necessarily involving forms of all degree. It would be interesting to study these theories, and characterise them in terms of the degrees of freedom they propagate as well as their dimensional reduction. We leave this to future work. 

Another interesting question is relation to geometry of spinors, especially in the nearly K\"ahler case. It is known, see \cite{Grunewald} and also \cite{Agricola:2014yma} that such manifolds admit Killing spinors. It would be interesting to find a characterisation of the field equations of the nearly K\"ahler case in spinor terms. Such a characterisation is known \cite{FKMU} in the 7D setting for the equation $dC={}^*C$ for a single 3-form in 7D. The 4-form ${}^*C$ is the Hodge dual of $C$ computed using the metric defined by $C$. In the 7D setup a 3-form is equivalent to a metric plus a unit spinor. Then the equation $dC={}^*C$ is known, see \cite{FKMU} to be equivalent to the Killing spinor equation for the corresponding spinor. 

\subsection{Organisation of the paper}

In Section \ref{sec:crit-points} we give some further characterisation of the critical points and in particular discuss how the metric is constructed in the nearly K\"ahler case. We perform the Hamiltonian analysis of all three theories in Section \ref{sec:Ham}, with the conclusion that the first two theories remain topological. We perform the dimensional reduction to 3D of all three theories in Section \ref{sec:dim-red}. We describe a particular solution to field equations in the nearly K\"ahler case in the Appendix.

\section{Characterisation of the critical points}
\label{sec:crit-points}

The aim of this section is to give some further characterisation of the critical points in each case. In particular, we describe why there is a naturally defined metric in the nearly K\"ahler case. 

\subsection{Symplectic manifolds}

We start by providing a characterisation of the critical points of theory (\ref{action-symp}), i.e. pairs $(B,C)$ satisfying 
\begin{equation}\label{feqs-symp}
dB=0, \qquad dC=- B\wedge B.
\end{equation}
As we have already explained, when the 2-form $B$ is non-degenerate, the first of the equations says that it is closed, and so is a symplectic form. So, critical points are symplectic manifolds.

There is however also the 3-form $C$ around, and one could try to use this additional structure to give a metric interpretation. However, as we shall now see, this does not seem to be possible in any natural way.

The construction one could attempt is as follows. As we shall explain below, a pair $B,C$ satisfying $B\wedge C=0$ naturally gives rise to a metric, which is obtained as $B(\cdot,J_C\cdot)$. We then notice that the theory (\ref{action-symp}) possesses the symmetry of shift of the 3-form $C$ by an exact form $C\to C+dH$, where $H\in \Lambda^2(M^6)$. This shift symmetry can then be used in an attempt to fund a preferred representative in the class of $C$'s related by adding an exact form. Thus, given a $C$, we will attempt to shift it by an exact form so that the desired property $B\wedge C$ holds
\begin{equation*}
    (C+dH)\wedge B=0.
\end{equation*}
It is, however, easy to see that there is no solution to this equation. Indeed, taking the exterior derivative and using the first of the equations (\ref{feqs-symp}) we have $dC\wedge B=0$. However, using the second equation we have $B\wedge B\wedge B=0$, which is a contradiction because we assumed that $B$ is non-degenerate. So, there appears to be no natural metric interpretation in this case. 

Another possible way to obtain $C$ with the desired property $B\wedge C=0$ is to simply shift $C$ by the wedge product of $B$ with some one-form $\theta$. It can be seen, see e.g. \cite{Krasnov:2017uam} section 2.2, that the 3-form $\tilde{C}=C+\theta\wedge B$ satisfying $\tilde{C}\wedge B=0$ is uniquely determined by $B,C$, provided some non-degeneracy assumption on these forms. Thus, one can always consider the metric obtained as$B(\cdot,J_{\tilde{C}} \cdot)$, but it is not clear what geometric significance this metric has, if any. 

\subsection{Complex manifolds}
We now come to the theory \eqref{action-complex} , the associated field equations are
\begin{equation}\label{feqs-complex}
dC=0,  \qquad dB = \hat{C}.
\end{equation}
As we have already explained, when $C$ is of the type that defines an almost complex structure,  (\ref{feqs-complex}) imply that $J_C$ is integrable. As in the previous case, there is the topological shift symmetry in this theory, which in this case is $B\mapsto B+d\theta$. As in the previous example, one could try to define a metric from the pair $B,C$. To do this we need $B\wedge C=0$, and so one could attempt to deform $B$ using the shift symmetry to satisfy this condition. However, this is not possible, for a similar reason as in the previous case. 

Indeed, the equation we want to impose is
\begin{equation*}
    (B+d\theta)\wedge C=0.
\end{equation*}
This is a set of six first order differential equations for six unknowns - the components of 1-form $\theta$. But unfortunately, there is no solution to this set of equations. Indeed, because $dC=0$ we can write this equation as $B\wedge C= - d(\theta\wedge C)$. We can then take the exterior derivative of both sides, and conclude that $0= dB\wedge C$. On the other hand, using the equation $dB=\hat{C}$ we reach a contradiction because $\hat{C}\wedge C\not=0$. The shift symmetry is thus not enough to obtain a 2-form $B$ with the desired property to define a metric. As in the previous case, one could instead decide to shift $C$ by $\theta\wedge B$. This is possible, but the geometrical interpretation, if any, is unclear. So, it appears that the critical points in this case are just complex manifolds, there is no natural metric interpretation. 

A metric interpretation will arise when we reduce this theory to 3 dimensions by compactifying on $S^3$. There is a metric on the $S^3$ fibers in this case, and so there is a well-defined inner product on vertical vector fields. Furthermore, having the complex structure at hand allows to define a notion of horizontal vector fields -- these are defined to be the $J_C$ images of the vertical ones. There is then a metric on the horizontal tangent space directly obtained by pulling back the vertical one. Practically, for any pair of horizontal vector fields, one first applies $J_C$ and then pairs the resulting vertical vector fields using the metric in the fibers. When $J_C$ is taken to have suitable $SU(2)$ invariance this metric then descends to a metric on the 3 dimensional base manifold, see \cite{Herfray:2016std} section 6.2 for more details. 

\subsection{Metric from a pair $B,C$}\label{subsection: Metric from B,C}

As we have already explained in the Introduction, the field equations in the nearly K\"ahler  case 
\be\label{feqs-nK}
    dC = - B\wedge B, \qquad dB = \hat{C}
\ee
imply $B\wedge \hat{C}=0$, and this is just the right condition to allow to construct a natural metric from the pair $B,\hat{C}$. Our goal now is to review this construction. We follow a description in \cite{Krasnov:2017uam}.

Recall \cite{Hitchin:2000jd} that a 3-form $C$ in 6 dimensions defines an endomorphism of the tangent bundle that squares to plus or minus identity. The sign depends on the ${\rm GL}(6,\R)$ orbit to which the 3-form belongs - there are exactly two orbits distinguished by this sign. 

We now recall how to construct this endomorphism. It will be convenient to choose some volume form $v$ on $M^6$. The end result however will only depend on the orientation of $v$. We first define an endomorphism $K_C$ that squares to a multiple of the identity, and then rescale. Let us define the action of $K_C$ on $\eta\in \Lambda^1(M^6)$ as follows
\be
i_\xi K_C(\eta) := \eta\, i_\xi C \wedge C/ v.
\ee
We emphasise that an arbitrary top form $v$ can be used in the denominator on the right-hand-side. On the left $K_C(\eta)$ is the 1-form that is the result of the action of $K_C$ on $\eta$. It can be verified that $K_C^2 = \lambda_C \id$ and so ${\rm Tr}(K_C^2) = 6\lambda_C$. It is convenient to define
\be
{\rm Vol}_C := \sqrt{\frac{\pm{\rm Tr}(K_C^2)}{6}} v.
\ee
Note that this is a well-defined volume form that depends only on the orientation of the auxiliary volume form $v$ used in the definition. We then define
\be
J_C := \frac{v}{{\rm Vol}_C} K_C.
\ee
The endomorphism $J_C$ depend on the volume form used in the construction of $K_C$ only via the induced choice of orientation. It squares to plus or minus identity, according to the sign of $\lambda_C$. Again this sign just characterizes the orbit to which the stable 3-form $C$ belongs. As a consequence, the resulting linear operator $J_C$ is either an almost complex structure, when $J_C^2=-\id$, or an almost para-complex structure, when $J_C^2=\id$. For definiteness, we shall assume from now-on that we are in the case of an almost complex structure.

If, in addition to $C$, we also have a 2-form $B$ satisfying $B\wedge C=0$, we can define a metric. In fact if $J_C$ is an almost complex structure, we even have an almost hermitian structure i.e a compatible triplet of 2-form, metric and almost complex structure. This comes from the following general remark (cf \cite{Hitchin:2000jd} ): any stable 3-form in 6 dimension lying in the `negative orbit' (i.e defining a complex structure) can be written $C = \gO + \overline{\gO}$ where $\gO$ is a $(3,0)$-form. Then $B \W C = 0$ implies that $B$ is a $(1,1)$-form. This in turns means that $B(J_C(X),J_C(Y))=B(X,Y)$ or equivalently $B(J_C(X),Y)=B(J_C(Y),X)$. Thus 
\be\label{metric-B}
g_B(X,Y) := B(J_C(X),Y).
\ee
is a good metric and $\left(g_B,B,J_{C}\right)$ is our almost Hermitian structure.

Now, in the positive orbit case, there is also a canonical decomposition $C = \ga^1\W\ga^2\W\ga^3 + \gb^1\W\gb^2\W\gb^3$ with $\ga^i$, $\gb^i$ eigenvectors of $\J_C$ with respectively eigenvalues $+1$ and $-1$. A reasoning along the same line as above then shows that $B \W C$ again implies that the tensor $g_B(X,Y) = B(J_C(X),Y)$ is symmetric.

All this is not surprising since $C,B$ can be thought of as components of a 3-form $\Omega= B\wedge dt +C$ in one dimension higher, and a 3-form in 7 dimensions defines a metric. However, as we just showed, the 6D metric induced by that 7D metric can be understood in purely 6D terms. 

We also give a direct verification of the fact that $B(J_C(X),Y)$ is symmetric when $B\wedge C=0$. Indeed, we have
\be
2B(X,J_C(Y)) = i_Y J_C(i_X B) = i_X B i_Y C\wedge C / {\rm Vol}_C.
\ee
We then use
\be
0= i_X (B \wedge i_Y C\wedge C) = i_X B \wedge i_Y C \wedge C + B \wedge i_X i_Y C\wedge C + B \wedge i_Y C \wedge i_X C
\ee
to see that when $B\wedge C=0$ we have
\be
2B(X,J_C(Y)) = - B i_X C\wedge i_Y C/ {\rm Vol}_C,
\ee
which is explicitly $X,Y$ symmetric. 

In the nearly K\"ahler case we have a pair $B,\hat{C}$ satisfying $B\wedge \hat{C}=0$, and so we can apply the above construction of the metric using $J_{\hat{C}}$. Note that $J_{\hat{C}}=J_C$.

\section{Hamiltonian analysis}
\label{sec:Ham}

The purpose of this section is to perform the Hamiltonian analysis of all 3 theories we introduced above. Our main goal is to verify that in the first two cases there are no propagating DOF, so the first two theories are topological.

We view $M^6 = \R \times M^5$. Let $t$ be a coordinate in the $\R$ direction. The 2- and 3-forms can be written as
\be\label{5+1}
B = dt \wedge \beta + b, \qquad C=dt\wedge \gamma + c.
\ee
Here $\beta\in \Lambda^1(M^5), b,\gamma\in \Lambda^2(M^5)$ and $c\in \Lambda^3(M^5)$. We will write the action in the Hamiltonian form for the most general case (\ref{action-nK}). It will be easy to get the actions for (\ref{action-symp}) and (\ref{action-complex}) by setting some of the terms to zero. Modulo a total derivative term we have
\be\label{action-Ham-form}
S[\beta,b,\gamma,c]=\int dt\int_{M^5} b\wedge \dot{c}  + \gamma\wedge db + \beta \wedge dc + {\rm vol}(\gamma,c) + \beta \wedge b\wedge b,
\ee
where ${\rm vol}(\gamma,c)$ is the result of evaluation of the second term in (\ref{action-nK}), and is an algebraic function of $\gamma,c$. The last term is the evaluation of the last term in (\ref{action-nK}). The "dot" stands for the time derivative. 

\subsection{Unreduced phase space, evolution equations and constraints}

In 5 dimensions 2- and 3-forms are dual to each other. So, from (\ref{action-Ham-form}) we see that the unreduced phase space is the space of pairs $(b,c)$ canonically conjugate to each other. The dimension of the unreduced phase space is twice the dimension of the space of 2-forms in 5D which is 10, so the phase space dimension is 20. 

Varying (\ref{action-Ham-form}) with respect to $b,c$ gives evolution equations, while varying with respect to $\beta,\gamma$ gives equations not containing time derivatives, i.e. constraints. For our purposes it is important to understand what these constraints imply. Varying with respect to $\beta$ we get the condition 
\be\label{constr-1}
dc=-b\wedge b.
\ee
In the case of the theory (\ref{action-complex}) we have zero on the right-hand-side of this equation instead, so the 3-form $c$ is closed in that case. 

Varying with respect to $\gamma$ gives a more involved equation because of the volume term. We get
\be\label{constr-2}
db + \frac{\partial {\rm vol}(\gamma,c)}{\partial \gamma} =0,
\ee
where the partial derivative of the volume 5-form with respect to the 2-form $\gamma$ is a 3-form. We will later see that some components of this equation have to be interpreted as constraints on $b$, while some other components give equations to (partially) determine $\gamma$ in terms of $db$. In the case of theory (\ref{action-symp}) this equation simply says $db=0$. 

The evolution equations are as follows
\be
\dot{c}=d\gamma-2\beta\wedge b, \qquad \dot{b} + d\beta = \frac{\partial {\rm vol}(\gamma,c)}{\partial c}.
\ee
In the case of theory (\ref{action-complex}) there is no second term on the right-hand side of the first equation. In the case of the theory (\ref{action-symp}) there is zero on the right-hand-side of the second equation. 

\subsection{Geometry of 3-forms on a 6 manifold with a time foliation}

To analyse the constraints, we will need to compute the volume term ${\rm vol}(\gamma,c)$ explicitly. The most efficient way of doing this is using some information about the endomorphism that $C$ defines. So, for concreteness, we will now assume that $C$ is of "negative type" i.e $\J_C$ is an almost complex structure on $M^6$. Everything we say can also be done for the other sign of $C$, with appropriate sign changes in the formulas. However, these sign changes will not affect the conclusion that the theories we study are topological, so it's enough to concentrate on the negative sign case.

Making a choice of time-foliation $t \from M^6=\R \times M^5 \to \R$ gives us a preferred one-form $dt$. The kernel of this one form defines the tangent space to $M^5$. We can act on $dt$ with $\J_C$ to obtain another one form:
\[ 
\tht = \J_C \lrcorner dt.
 \]
It is an easy calculation to check that 
\[ 
A= dt +i \tht
 \]
is $(0,1)$ for $\J_C$, i.e $J_C \lrcorner A = -i A$. From this it follows that the distribution $D^4_C$ defined by
\[ 
D^4_C = Ker\left(dt\right) \cap Ker\left(\tht \right) = Ker\left(A\right),
 \]
it stabilised by $\J_C$. In particular $\J_C$ restricted to $D^4_C$ gives an almost complex structure on this distribution.

Another useful construction for later purposes is as follows. Let us pick a particular set of "spatial" coordinates $x^a$ on $M^5$.  We can then define a vector field $\pa_t$ "normal to $M^5$" in $M^6$. Practically this vector fiels is such that its insertion into all one-forms $dx^a$ is zero. Acting on this vector with $\J_C$ we obtain another vector
\[ 
N := \J_C\left(\pa_t\right).
 \]
Both $N$ and $\pa_t$ are "normals to $D^4_C$ in $M^6$". That $N$ is not in $D^4_C$ is readily seen from
\[ 
\tht(N) = \left(\J_C\right)^2(\pa_t) \lrcorner dt = -dt(\pa_t)= -1
 \]
 To avoid any misconception at this point it is important to point out that, generically, $N$ does not lie in $M^5$,
	\[ dt(N) = \tht(\pa_t) \neq 0. \] 
Note that while the distribution $D^4_C$ depends only on the foliation of $M^6$ by hypersurfaces $M^5$, the "normal" vectors $\pa_t$ and $N$ depend on the specific choice of spatial coordinates on $M^5$ and are thus less canonical.

\subsection{Computation of the volume}

Following Hitchin we introduce the (densitised) endomorphism of $T^* M^6$
\begin{IEEEeqnarray*}{lCr}
	\Jt_C&=& \frac{1}{6} \epst^{\mu \ga\gb\gc\gd\eps} C_{\ga\gb\gc} C_{\gd\eps\nu} \;dx^{\nu}\otimes \pa_{\mu}.
\end{IEEEeqnarray*}
We then use the decomposition $C = dt \W \gc + c$ and coordinates $\left(x^{\mu}\right) = \left(t , x^a \right)$. A computation gives
\begin{IEEEeqnarray*}{lCr}
	\Jt_C &\quad =\quad & (\tilde{c}^{ab}\gamma_{ab}) \left( dt \otimes \pa_t - \delta^b_a dx^a \otimes \partial_b\right) + \tilde{\theta}_a dx^a \otimes \partial_t 
	+  dt \otimes \Nt^b \partial_b + 4 \tilde{c}^{bc}\gamma_{ac} \;dx^a\otimes \pa_b.
\end{IEEEeqnarray*}
Here we used the fact that a 3-form $c\in \Lambda^3(M^5)$ is dual to a densitised bivector 
\be
\tilde{c}^{ab}:=\frac{1}{6} \tilde{\epsilon}^{abcde} c_{cde}, \qquad 
c_{abc} = \frac{1}{2}\epsut_{abcde}\ct^{de} .
\ee
The objects $\tilde{\epsilon}^{abcde}, \epsut_{abcde}$ are densitiesed anti-symmetric tensors that exist on $M^5$ without any metric. The other objects are as follows
\begin{IEEEeqnarray}{lCr C lCr C lCr}
\thtt_a &:=& \ct^{bc}c_{abc}  &,\qquad & \Nt^a &:=& -\frac{1}{2} \epst^{abcde} \gc_{bc} \gc_{de} &.
\end{IEEEeqnarray}
It is easy to check that $\Jt_C$ is tracefree, as it should be.

To compare to our previous discussion, we have introduced $\thtt = \Jt_C \lrcorner dt$ and $\Nt = \Jt_C(\pa_t)$. These objects compute to
\be
\thtt = (\tilde{c}^{ab}\gamma_{ab}) dt + \tilde{\theta}_a dx^a, \qquad \Nt= (\tilde{c}^{ab}\gamma_{ab}) \partial_t + \Nt^a \partial_a.
\ee
So, $\thtt_a, \Nt^a$ are indeed the "spatial" parts of the one-form $\thtt$ and vector field $\Nt$, which justifies the notation. 

We remark that $\tilde{\theta}_a$ is in the kernel of $\tilde{c}^{ab}$. Indeed, we have
\be
\ct^{ab} \tilde{\theta}_b= \ct^{ab} \ct^{cd} c_{bcd} = \frac{1}{2}\epsut_{bcdef}\ct^{ab} \ct^{cd}\ct^{ef}=0 
\ee
because for any anti-symmetric tensor $c^{ab}$ we have $c^{a[b} c^{cd} c^{ef]}=c^{[ab} c^{cd} c^{ef]}$, and anti-symmetrisation over 6 indices in 5 dimensions vanishes. 

Let us now compute the volume. A brute force way of doing this is to compute $\Jt_C^2$ and then take the trace. However, we can make use of our knowledge on $\Jt_{C}$ to shortcut this calculation: because its square is proportional to identity it is indeed enough to compute $\thtt(\Nt)= dt\left(\;\Jt^2(\pa_t)\right)$. This will immediately give us (minus) the volume squared, $\Jt_{C}^2 = -(Vol_{C})^2 \id$. We have
\be
\thtt(\Nt) = (\tilde{c}^{ab}\gamma_{ab})^2 - \frac{1}{4} \tilde{\epsilon}^{abcde} \gamma_{bc}\gamma_{de} 
\,\, \epsut_{ab'c'd'e'} \tilde{c}^{b'c'}\tilde{c}^{d'e'} \\ \nonumber
= 4 \tilde{c}^{ac} \gamma_{bc} \tilde{c}^{bd} \gamma_{ad} - (\tilde{c}^{ab}\gamma_{ab})^2.
\ee
From which it follows that
\be\label{vol}
{\rm vol}(\gamma,c) = \sqrt{(\tilde{c}^{ab}\gamma_{ab})^2- 4 \tilde{c}^{ac} \gamma_{bc} \tilde{c}^{bd} \gamma_{ad}}.
\ee
We have checked by an explicit computation that the same result is obtained by computing the trace of $\Jt_C^2$. We have also checked that for the canonical form of negative type $C={\rm Re}(\alpha_1\alpha_2\alpha_3)$ with $\alpha_i=dx_i+\im dy_i$, and say the choice $x_1\equiv t$, the expression under the square root is a positive constant, so that the square root is real. 

We note that a precise numerical constant in the formula for the volume is unimportant, for it contributes to the numerical coefficient in front of the second term in (\ref{action-complex}), and this can always be absorbed by redefining the fields $B,C$.

\subsection{Constraints}

Now that that we have the expression (\ref{vol}) for the volume we can write the constraint (\ref{constr-2}) more explicitly. We get
\be\label{constr-2*}
\widetilde{(db)}^{ab} + \frac{1}{{\rm vol}(\gamma,c)} M^{ab\, cd} \gamma_{cd} =0,
\ee
where $\widetilde{(db)}^{ab}=(1/6)\tilde{\epsilon}^{abcde} (db)_{cde}$ is the dual of the exterior derivative of $b$ and 
\be
M^{ab\, cd} =  \tilde{c}^{ab} \tilde{c}^{cd} + 4 \tilde{c}^{a[c} \tilde{c}^{d]b}
\ee
is a symmetric matrix (of density weight two) mapping two-forms into bivectors. 

It is useful to note that the norm of $\gc$ in the scalar product given by $M^{ab\, cd}$ is just ${\rm vol}(\gc, c)$. Indeed, we have
\be
|\gamma|^2_{M} \coloneqq \gc_{ab} \;M^{ab\,cd} \;\gc_{cd} = {\rm vol}(\gc, c) \;\gc_{ab} \frac{\pa {\rm vol}(\gc,c)}{\pa \gc_{ab}} = \left( {\rm vol}(\gc,c) \right)^2 
\ee
where we used that ${\rm vol}(\gc,c)$ is a homogeneous function of degree one in $\gamma$. Equation \eqref{constr-2*} can thus be rewritten 
\be\label{constr-2*bis}
\widetilde{(db)}^{ab} + M^{ab\, cd} \frac{\gamma_{cd}}{|\gamma|_{M}} =0.
\ee
This way of writing the constraint makes it clear that the overall scale of $\gamma_{ab}$ cannot be solved for from the constraint. This leads to the overall scale of $\gamma_{ab}$ to remain an arbitrary function in the evolution equations. Below the presence of such an arbitrary function will be shown to be related to the freedom of performing temporal diffeomorphisms. 

There are some more components of $\gamma_{ab}$ that cannot be solved from \eqref{constr-2*bis}.
It can be checked that for a generic $\tilde{c}^{ab}$ the matrix $\tilde{M}^{ab\, cd}$ in 5 dimensions has rank 5, and that its zero eigenvectors are of the form $\gamma^0_{ab} = (1/2) \epsut_{abcde} \xi^c \tilde{c}^{de}$ for an arbitrary vector field $\xi^a$. In other words, the zero directions of $\tilde{M}^{ab\, cd}$ are
\be\label{gamma-0}
\gamma^0 = i_\xi c,
\ee
where $i_\xi$ is the interior product. To check that these are indeed the zero eigenvectors we need the identity 
\be
\tilde{c}^{b[d} \tilde{c}^{ef]} = \tilde{c}^{[bd} \tilde{c}^{ef]} = \frac{1}{24} \tilde{\epsilon}^{bdefh}\epsut_{hmnpq} \tilde{c}^{mn} \tilde{c}^{pq}.
\ee
This identity is then used in the computation of the insertion of $\gamma^0_{cd}$ into the second term of the matrix
\be
- 4 \tilde{c}^{ac} \tilde{c}^{bd} \frac{1}{2} \epsut_{mcd ef} \xi^m \tilde{c}^{ef} = - \frac{1}{2} \tilde{c}^{ab}\epsut_{mcd ef} \xi^m \tilde{c}^{cd} \tilde{c}^{ef},
\ee
and so insertion into the second term cancels the insertion into the first term of $\tilde{M}^{ab\, cd}$.

Thus, the endomorphism $M$ as a non-trivial kernel and we learn that the components of $\gamma$ which are of the form $i_\xi c$ cannot be solved for from the constraint (\ref{constr-2*}). A related observation is that the image of $M$ is not the whole space of bivectors. Instead, the presence of a non-trivial kernel for $M$ implies that if $\tilde{\go}^{ab} = \epst^{abcde} \go_{cde}$ is in the image of $M$
\be \label{image of M}
\tilde{\go}^{ab} = M^{abcd} \gc_{cd}
\ee
it must satisfy some constraints. These are easily found by contracting the equation (\ref{image of M}) with $\gamma^0$ of the form (\ref{gamma-0}). We get
\be \label{constr image of M}
\go \W i_\xi c=0,\; \forall \xi.
\ee 
As a result we obtain the following set of $\gc$-independent constraints on $b$:
\be\label{constr*}
db \wedge i_\xi c=0.
\ee
This must hold for an arbitrary vector field $\xi$, and gives a set of 5 constraints on the unreduced phase space. We will soon see that these constraints generate diffeomorphisms of the spatial slice.

For purposes of analysing the constraint algebra we also need to rewrite the Hamiltonian constraint that follows from \eqref{constr-2*bis} as an equation that is $\gamma$-independent. Looking at \eqref{constr-2*bis}, one can interpret the constraint equations as follows: not only must $\widetilde{db}$ be in the image of $M$, i.e must verify \eqref{constr*}, but it should also be the image of a unit length vector. This second constraint can be written in an implicit form: Let $N_{ab\,cd}$ be an inverse of $M^{ab\; cd}$
\be
\gat^{ab} N_{ab\,cd} M^{cd\; ef} = \gat^{ef}, \qquad \forall\; \gat^{ab} \coloneqq \epst^{abcde} \ga_{cde}.
\ee
This inverse is only well defined on the image of $M$, i.e the preceding equation only make sense when $\ga\W i_{\xi}c=0$. Moreover, $\gat^{ab} N_{ab\,cd}$ is really an element of the quotient space of 2-forms modulo the kernel of $M$, i.e. is defined only up to $i_{\xi}c$. Since $M$ is only a function of $c$ so is $N$, when it is well defined.

Now \eqref{constr-2*bis} implies that $\widetilde{db}$ has unit length in the quadratic form defined by $N$:
\be
\widetilde{db}^{ab} \;N_{ab\; cd} \;\widetilde{db}^{cd} = \frac{\gamma_{ab}}{|\gamma|_{M}} M^{ab\; cd} \frac{\gamma_{cd}}{|\gamma|_{M}} = 1.
\ee
We thus can write the "Hamiltonian" constraint in the form
\be \label{constr**}
\widetilde{db}^{ab} \;N_{ab\; cd} \;\widetilde{db}^{cd} =1.
\ee

\subsection{Reduced phase space: symplectic case}

In the symplectic case we have the following constraints
\be
dc=-bb, \qquad db=0.
\ee
These constraints can be converted into generators of transformations on the unreduced phase space, by integrating them against appropriate smearing functions. So, we define
\be\label{Q-theta}
Q_\theta := \int \theta (dc+bb), \qquad Q_h := \int h db.
\ee
Here $\theta$ is a 1-form, and $h$ is a 2-form. The constraints $Q_h$ clearly commute with themselves, as well as with $Q_\theta$. The last assertion is seen from the fact that $\{Q_\theta, Q_h\}$ contains $d^2$ and thus vanishes. 

Let us also compute the commutator of two $Q_\theta$. We have
\be\label{comm-Q-theta}
\{ Q_{\theta_1}, Q_{\theta_2}\} = \{ \int d\theta_1 c + \theta_1 bb, \int d\theta_2 c + \theta_2 bb \,\} = \\ \nonumber
-2\int (d\theta_1 \theta_2 - d\theta_2 \theta_1) b = -2\int d(\theta_1 \theta_2) b = 2\int \theta_1\theta_2 db = 2 Q_{\theta_1\theta_2}.
\ee
Our Poisson bracket is $\{b(x),c(y)\}=\delta(x,y)$. We thus see that the algebra of the constraints closes. 

Let us now determine the reduced phase space. The constraints generate the following transformations
\be
\{ Q_\theta, b\} = - d\theta, \qquad \{ Q_h, c\} = -d h.
\ee
Thus, on the constraint surface $db=0$ modulo the action of $Q_\theta$ the space of $b$'s reduces to cohomology - closed forms modulo exact forms. For the purpose of counting the local degrees of freedom we can assume that the relevant cohomology is trivial. With this assumption, on the constraint surface $db=0$ modulo transformations generated by $Q_\theta$ we have $b=0$. 

With $b=0$ the other constraint becomes $dc=0$, and on this constraint surface modulo the action generated by $Q_h$ the space of $c$'s also reduces to the cohomology. So, the reduced phase space is at most finite dimensional and the theory is topological.

\subsection{Reduced phase space: complex case}

In this case the constraints are
\be\label{complex-constraints}
dc=0, \qquad i_\xi c db=0, \qquad \widetilde{db}^{ab} \;N_{ab\; cd} \;\widetilde{db}^{cd} =1.
\ee
First, let us check the algebra of constraints smeared with appropriate test functions. It is easy to see that because the first constraint involves $dc$ and the other two constraints involves $db$, the Poisson bracket involves $d^2$ and is thus zero. Thus, the first set of constraints commutes with all other constraints. 

However, the second set of constraints does not commute. In fact, it should not, because the expectation is that the second and third constraints are the generators of the spatial and temporal diffeomorphisms. Thus, they should form the usual algebra. Let us verify this.

To compute the algebra it is convenient to replace the second constraint by a linear combination with the first constraint. Thus, we can write the integrated version of the second constraint as
\be
Q_\xi :=-\int db \wedge i_\xi c = \int b \wedge di_\xi c = \int b \wedge {\cal L}_\xi c - \int b \wedge i_\xi dc.
\ee
Thus, modulo the first constraints the second set of constraints is equivalent to 
\be
Q'_\xi = \int b \wedge {\cal L}_\xi c,
\ee
where ${\cal L}_\xi= di_\xi + i_\xi d$ is the Lie derivative. It is then clear that what $Q'_\xi$ generates are just diffeomorphisms of the "spatial" slice $M^5$
\be
\{ Q'_\xi, c\} =  {\cal L}_\xi c, \qquad \{ Q'_\xi, b\} = {\cal L}_\xi b,
\ee
The algebra of these constraints is then obviously the algebra of diffeomoprhisms 
\be
\{ Q'_{\xi_1}, Q'_{\xi_2}\} = -\int b\wedge \left( {\cal L}_{\xi_1}{\cal L}_{\xi_2} -{\cal L}_{\xi_2}{\cal L}_{\xi_1}\right) c \equiv - Q'_{[\xi_1,\xi_2]}  .
\ee
Similarly, the Poisson bracket of the spatial diffeomorphism constraint $Q'_\xi$ with the last constraint in (\ref{complex-constraints}) is again the last constraint with the Lie derived smearing function. 

We now consider the smeared version of the last constraint in \eqref{complex-constraints}
\be
Q_{\gl} = \int \gl \widetilde{db}N\widetilde{db},
\ee
where we dropped an unimportant constant piece. Here $N\widetilde{db}$ is a schematic notation for the 3-form $N_{ab\;cd} \widetilde{db}^{cd} dx^a\W dx^b$. We need to compute the Poisson bracket of this constraint with itself, smeared with different test functions. The expectation is that the result is a spatial diffeomorphism constraint.

Instead of doing this calculation explicitly, which seems hard, let us verify that the transformation this constraint generates indeed corresponds to a temporal diffeomorphism. The infinitesimal gauge transformation of $c$ generated by $Q_{\gl}$ is
\be \label{c gauge transform}
\left\{Q_\gl , c \right\} = d \left( \gl N \widetilde{db} \right).
\ee
Note that the 2-form $\gl N \widetilde{db}$ is only defined up to $i_{x}c$, as a result of the ambiguity in the definition of N, but that this only adds an infinitesimal spatial diffeomorphism to this transformation. On the other hand, to find the result of a temporal diffeomorphism we take the original 6-dimensional 3-form $C=dt \gamma + c$ and consider 
\be
{\cal L}_{\lambda \partial/\partial t} C = d( \lambda \gamma) + \lambda (\dot{c}-d\gamma).
\ee
But our evolution equation is $\dot{c}=d\gamma$ and so the second term drops. On the other hand, $N \widetilde{db}$ in \eqref{c gauge transform} is just a multiple of $\gamma$, as follows from the definition of $N$. So, \eqref{c gauge transform} is indeed an infinitesimal temporal diffeomorphisms, as we expected, and therefore the commutator of two such transformations is a spatial diffeomorphism. This shows that the algebra of constraints closes. 

Let us now analyse the reduced phase space. First, we analyse the space of $c$'s. We have the constraint $dc=0$, and in addition have to mod out by the action generated by the constraints. The action generated by $Q'_\xi$ on the surface $dc=0$ is $\delta c = d(i_\xi c)$. The action of the temporal diffeomorphism is \eqref{c gauge transform}. Now, on the surface $dc=0$ the 3-form $c$ is exact, modulo potentially present harmonic pieces that are of no interest to us, as we just want to verify that there are no local degrees of freedom. The space of exact 3-forms has the dimension of the space of 2-forms, modulo one forms, which are in turn taken modulo zero forms. Thus, the dimension is $10-(5-1)=6$. On the other hand, we have precisely 6 gauge transformation parameters available to us: 5 in the spatial diffeomorphisms and an additional one in the temporal diffeomorphisms. Thus, it should be possible to set the exact part of $c$ to zero (modulo harmonic forms) by these transformations.

So, we need to show that the space of 2-forms can be written as a direct sum of exact forms and the span of $i_{\xi} c$ and $\gl N \widetilde{db}$. These subspace respectively are four and six dimensional, thus we just need to show that they do not intersect. However, generically $c$ has no killing vectors and $\Lc_{\xi} c = d(i_{\xi} c) \neq 0 \forall \xi$. So generically $i_{\xi} c$ is not closed and cannot be exact. On the other hand, rewriting the Hamiltonian constraint as
\be
0= \int \left(\gl N \widetilde{db}\right) \W db +\gl =  \int -d\left(\gl N \widetilde{db}\right) \W b +\gl,
\ee
we see that $\gl N \widetilde{db}$ cannot be closed everywhere and is thus not exact.

The analysis for $b$ is similar. We have already used diffeomorphisms to get rid of the degrees of freedom contained in $c$. All the gauge transformation we are left with is shifting of $b$ by exact forms. The space of 2-forms can be parametrised as forms that are co-exact, i.e. $d^* b_3$, plus exact forms $db_1$ which are pure gauge, plus possibly harmonic forms which we ignore. The first space is the space of 3-forms modulo 4-forms in turn taken modulo 5-form, which gives the dimension $10-(5-1)=6$. We have in our disposal 6 constraints to kill the 6-dimensional co-exact part of $b$, which is just the number that is needed. Thus the reduced phase space is at most finite dimensional, and there are no local degrees of freedom in this version of the theory. 

Another verification of the absence of the degrees of freedom in this theory is the explicit computation of the one-loop effective action performed in \cite{Pestun:2005rp}. This paper considered a slightly different theory from ours, namely the action used was just the volume term, but restricted to the space of closed 3-forms. It was then shown that the one-loop effective action is a certain ratio of two independent analytic torsions that exist on a complex 3-manifold. This analysis is applicable to our case as well, as the role of the first term in the action (\ref{action-complex}) is just to set $C$ to be closed. 

\subsection{Nearly K\"ahler case}

In the nearly K\"ahler case the constraints are
\be\label{nK-constraints}
dc=-bb, \qquad i_\xi c db=0, \qquad \widetilde{db}^{ab} \;N_{ab\; cd} \;\widetilde{db}^{cd} =1.
\ee
We have already computed the Poisson brackets of the first of these constraints (\ref{Q-theta}) with itself, see (\ref{comm-Q-theta}). The result of this Poisson bracket was a different constraint $db=0$. In the symplectic case above this constraint was already present and the algebra closed. In the current case, however, we get new constraint $\theta_1\theta_2 db=0$, which says that $db$ smeared with an arbitrary {\it simple} 2-form is zero. This constraint has some intersection with the other constraints present in (\ref{nK-constraints}), but does not coincide with them. So, we need to add the new constraints with new Lagrange multipliers into the action, and keep applying the Dirac procedure. We take the fact that the algebra of the constraints does not close at the first step as a hint indicating that there are propagating degrees of freedom on this case. But we have not completed the Hamiltonian analysis in this much more involved case, so we refrain from making any statement about the nature of the theory in this case. 

One reason that makes us believe that there are propagating degrees of freedom is similarity of the action (\ref{action-nK}) to the Hamiltonian form of the action of the theory analysed in \cite{Krasnov:2017uam}. In the latter case, making an assumption that all fields are $t$-independent, one obtains a theory with the kinetic term $BdC$ but with a potential term different from the one in (\ref{action-nK}), and instead involving a product of powers of $BBB$ and $C\hat{C}$. This 7-dimensional theory is known to have propagating degrees of freedom, see \cite{Krasnov:2017uam}. Degrees of freedom do not disappear in the process of dimensional reduction. So, we have an example of a 6D theory with kinetic term $BdC$ and a potential term depending on both $B$ and $C$, but in a way more involved than in (\ref{action-nK}). This theory does describe propagating degrees of freedom, and this makes us suspect that this is also the case for the theory (\ref{action-nK}), but we will leave this question to future research. 

\section{Dimensional reduction}
\label{sec:dim-red}

We now carry out the dimensional reduction to 3D of all 3 theories described above. We make an assumption that we have the group ${\rm SU}(2)$ acting on $M^6$ without fixed points, which gives $M^6$ the structure of the principal ${\rm SU}(2)$ bundle over a 3-dimensional base. We assume that the 3- and 2-forms are invariant under this action, and parametrise them by data on the base. This is similar to what was done in \cite{Krasnov:2017uam}, the only difference being that the base is now 3-dimensional. 

\subsection{Parametrisation}

We choose to parametrise the 3-form in the following way
\be\label{C}
C = -2 {\rm Tr}\left( \frac{1}{3} \phi^3\, W^3 - \phi W E^2\right) + c.
\ee
Here $\phi$ is a scalar field, $c$ is a 3-form on the base and 
\be\label{lift}
W= g^{-1} dg+A, \qquad A = g^{-1} {\bf a} g, \qquad E=g^{-1} {\bf e} g
\ee
are a connection on the total space of the bundle and the lifts of the Lie algebra valued 1-form $\bf a$ and 1-form $\bf e$ on the base to the total space of the bundle. To write the second term of the parametrisation (\ref{C}) we have used the fact that any Lie algebra valued 2-form field on the 3-dimensional base is, up to the sign, the wedge product of two Lie algebra valued 1-forms. We have chosen the sign that corresponds to 3-forms of negative type, which is the most interesting case. A simple computation then gives
\be
dC= - 2 \,{\rm Tr} \Big( \phi^2 d\phi W^3 +( \phi^3 F- \phi E^2) W^2 \\ \nonumber 
-d_A(\phi E^2) W - \phi FE^2 \big).
\ee
Here $F=g^{-1} {\bf f} g$ is the lift to the total space of the curvature ${\bf f}=d{\bf a}+{\bf a}^2$, and $d_A E = g^{-1} (d{\bf e}+{\bf a}{\bf e}+{\bf e}{\bf a}) g$ is the lift to the bundle of the covariant derivative of Lie algebra-valued 1-form ${\bf e}$ with respect to the connection $\bf a$. There is no term $dc$ on the right-hand-side in the above formula because the base is 3-dimensional. 

Similarly, we parametrise the 2-form $B$ as follows
\be\label{B-reduction}
B= - 2{\rm Tr} \left( \Psi W^2 - \Theta W\right) + b.
\ee
Here $\Psi=g^{-1} \bm{\psi} g$ is the lift to the total space of the bundle of a Lie algebra valued scalar field on the base, and $\Theta= g^{-1} \bm{\theta} g$ is the lift of a Lie algebra valued 1-form field. The object $b$ is a two-form field on the base. 

\subsection{Dimensionally reduced action}

We now compute the dimensional reduction of the pieces of the action. For the kinetic term we get
\be
BdC / \left( -\frac{2}{3}{\rm Tr} (W^3)\right) = - 2{\rm Tr}\left( \bm{\psi} d_{\bf a} (\phi {\bf e}^2) + \bm{\theta} (\phi^3 {\bf f} - \phi {\bf e}^2) \right) + b d(\phi^3) ,
\ee
where we have divided by the volume form in the fiber. 

For the $B^3$ term we have
\be
\frac{1}{3} B^3 / \left( -\frac{2}{3}{\rm Tr} (W^3)\right)  = - 2{\rm Tr}\left( 2 \bm{\psi} \bm{\theta} b - \frac{2}{3} \bm{\theta}^3\right).
\ee

In deriving this result we have used the following trace identities
\be
(-2{\rm Tr}(W^2 a))(-2{\rm Tr}(W b)) = -\frac{2}{3} {\rm Tr}(W^3) ( -2{\rm Tr}(ab)), \\ \nonumber
(-2{\rm Tr}(W a))(-2{\rm Tr}(W b)) (-2{\rm Tr}(W c))=-\frac{2}{3} {\rm Tr}(W^3) ( -4{\rm Tr}(abc)).
\ee

To compute the volume term we parametrise 
\be
c = \frac{4}{3} \rho {\rm Tr}(E^3)
\ee
for some function $\rho$ on the base. The volume term is then computed in two steps. First, one computes the volume for $\rho=0$. This is done by noting that 
\be
C= - \frac{2}{3}{\rm Re} \left({\rm Tr}(A^3)\right), \qquad A= W+ \im E.
\ee
This immediately implies that 
\be
\hat{C} = - \frac{2}{3}{\rm Im} \left({\rm Tr}(A^3)\right) = - 2{\rm Tr}\left( \phi^2 W^2 E - \frac{1}{3} E^3\right).
\ee
Then we get
\be
{\rm vol}_C = \frac{1}{2} C \hat{C} = -\frac{2}{3}{\rm Tr}(W^3) \frac{\phi^3}{3} {\rm Tr}(E^3).
\ee

The dependence on $\rho$ has been computed in \cite{Herfray:2016std}, formula (159). As is shown in this reference, one just needs to multiply the $\rho=0$ result by $\sqrt{1-\rho^2}$. Thus, overall we get
\be
{\rm vol}_C / \left( -\frac{2}{3}{\rm Tr} (W^3)\right) = -2{\rm Tr}\left( -\frac{\phi^3}{6} E^3 \sqrt{1-\rho^2}\right).
\ee

\subsection{Symplectic case}

In the case of theory (\ref{action-symp}) the dimensionally reduced action becomes
\be
S_{symp}[\phi,{\bf a},{\bf e},\bm{\psi},\bm{\theta},b]=\int -2{\rm Tr}\left( \bm{\psi} d_{\bf a} (\phi {\bf e}^2) + \bm{\theta} (\phi^3 {\bf f} - \phi {\bf e}^2)  - 2\bm{\psi} \bm{\theta} b + \frac{2}{3} \bm{\theta}^3 \right) + b d(\phi^3).
\ee

Let us write down the Euler-Lagrange equations that follow by extremising this action. Varying with respect to $b$ we get
\be\label{eqn-symp-1}
d(\phi^3) + 4{\rm Tr}(\bm{\psi} \bm{\theta}) =0.
\ee
Varying with respect to $\bm{\psi}$ we get
\be\label{eqn-symp-2}
d_{\bf a}(\phi {\bf e}^2) = 2\bm{\theta} b.
\ee
Varying with respect to $\bm{\theta}$ gives
\be\label{eqn-symp-3}
\phi {\bf e}^2 - \phi^3 {\bf f} + 2\bm{\psi} b= 2\bm{\theta}^2.
\ee
Varying with respect to ${\bf e}$ gives
\be\label{eqn-symp-4}
d_{\bf a}\bm{\psi} {\bf e} + {\bf e} d_{\bf a}\bm{\psi} + \bm{\theta} {\bf e} + {\bf e} \bm{\theta} =0.
\ee
Varying with respect to $\phi$ gives
\be\label{eqn-symp-5}
2{\rm Tr}\left( d_{\bf a}\bm{\psi} \phi {\bf e}^2  + \bm{\theta} \phi {\bf e}^2 - 3\phi^3 \bm{\theta} {\bf f} \right) = 3\phi^3 db.
\ee
Finally, varying with respect to the connection we get
\be\label{eqn-symp-6}
\phi(\bm{\psi} {\bf e}^2 - {\bf e}^2 \bm{\psi}) = d_{\bf a} (\phi^3 \bm{\theta}). 
\ee

To analyse these equations it is most convenient to take advantage of the gauge symmetry of this theory from the start. The original 6D theory is invariant under $C\to C+dH$, where $H$ is a 2-form. It can then be seen that this shift is in particular equivalent to shifts of the connection $a\to a+\xi$. This gives rise to the following symmetry of the above action, in infinitesimal form
\be
\delta a = \chi, \quad \delta (\phi {\bf e}^2) = d_{\bf a} (\phi^3 \chi), \quad \delta \bm{\theta} = \bm{\psi} \chi - \chi \bm{\psi}, \quad \delta b = -2{\rm Tr}(\bm{\theta} \chi).
\ee
This gauge symmetry can be used to set $a=0$ from the beginning, which is convenient to do. 

We can then use (\ref{eqn-symp-4}) in (\ref{eqn-symp-5}). Assuming $\phi\not=0$, we get
\be\label{symp-b}
db = 0.
\ee
All other equations become as follows
\be\nonumber
d(\phi^3) + 4{\rm Tr}(\bm{\psi} \bm{\theta}) =0, \\ \nonumber
d(\phi {\bf e}^2) = 2\bm{\theta} b, \\ 
\phi {\bf e}^2 + 2\bm{\psi} b= 2\bm{\theta}^2, \\ \nonumber
d\bm{\psi} {\bf e} + {\bf e} d\bm{\psi} + \bm{\theta} {\bf e} + {\bf e} \bm{\theta} =0, \\ \nonumber
\phi(\bm{\psi} {\bf e}^2 - {\bf e}^2 \bm{\psi}) = d(\phi^3 \bm{\theta}). 
\ee
It would be interesting to find non-trivial examples of solutions to this system. 

\subsection{Complex case}

This is the most interesting case for us, because of expectation that the dimensional reduction in this case gives 3D gravity. 

In the case (\ref{action-complex}) the dimensionally reduced action is
\be\label{action-red-compl}
S_{compl}[\phi,{\bf a},{\bf e},\rho,\bm{\psi},\bm{\theta},b]=\int -2{\rm Tr}\left( \bm{\psi} d_{\bf a} (\phi {\bf e}^2) + \bm{\theta} (\phi^3 {\bf f} - \phi {\bf e}^2)  -\frac{\phi^3}{6} \sqrt{1-\rho^2} \, {\bf e}^3 \right) + b d(\phi^3).
\ee
Varying with respect to $\rho$ we immediately get $\rho=0$, and so we can set $\rho$ to zero in what follows. Varying with respect to $b$ we get that $\phi=const$. Varying with respect to $\bm{\psi}$ we get $d_{\bf a} {\bf e}=0$. Varying with respect to $\bm{\theta}$ gives 
\be\label{const-curv}
\phi^2 {\bf f} = {\bf e}^2.
\ee
Thus, we already get the correct field equations of 3D gravity with non-zero (negative in this case) cosmological constant. The appearance of a particular sign for the cosmological constant is related to our choice to restrict ourself to a particular orbit (the ``negative one") for 3-form. 

Varying with respect to ${\bf e}$ gives
\be\label{eqs-complex-1}
d_{\bf a} \bm{\psi}  + \bm{\theta} + \frac{\phi^2}{4} {\bf e} =0,
\ee
where we assumed that $\phi\not=0$ and that ${\bf e}$ is non-degenerate so that it can be cancelled from this equation. Varying with respect to ${\bf a}$ we get
\be\label{eqs-complex-2}
\bm{\psi} {\bf e}^2 + {\bf e}^2 \bm{\psi} + \phi^2 d_{\bf a} \bm{\theta}=0.
\ee
Finally, varying the action with respect to $\phi$ one gets an equation for $db$ that we do not care about because $b$ is an auxiliary field. 

We can now use the gauge symmetry of this system to simplify the equations. Thus, the original action is invariant under shifts $B\to B+ dX$ for some 1-form $X$. This shift symmetry can be used to kill the $\bm{\psi}$ component of $B$ from the very beginning. Indeed, taking $X=2{\rm Tr}(\Psi W)$ gives 
\be
dX = - 2{\rm Tr}\left( \Psi W^2 - d_A \Psi W - \Psi F\right).
\ee
Thus, we can kill $\bm{\psi}$ at the expense of shifting $\bm{\theta}$ and $b$ in (\ref{B-reduction}). Thus, it is convenient to set $\bm{\psi}$ to zero using this shift symmetry. 

Setting $\bm{\psi}=0$ in (\ref{eqs-complex-1}) gives $\bm{\theta}= -(\phi^2/4){\bf e}$, which in turn makes (\ref{eqs-complex-2}) satisfied because of $d_{\bf a} {\bf e}=0$. So, we get the correct equations of 3D gravity with negative cosmological constant. 

Overall, we get the field equations of 3D gravity by the dimensional reduction, as could have been expected from the fact that the 6D almost complex structure in this case is integrable. Setting from the beginning $\bm{\psi}=0$ and $\phi=const$, the dimensionally reduced action (\ref{action-red-compl}) is essentially that of the 3D gravity, apart from the fact that there is an additional 1-form field $\bm{\theta}$ in it. Varying this action with respect to $\bm{\theta}$ then gives the constant curvature condition (\ref{const-curv}), varying with respect to ${\bf e}$ identifies $\bm{\theta}$ with a multiple of ${\bf e}$, and varying with respect to the connection gives $d_{\bf a}\bm{\theta}=0$ which implies $d_{\bf a}{\bf e}=0$. So, we get an acceptable Lagrangian formulation of 3D gravity with non-zero cosmological constant if we from the beginning set $\bm{\psi}=0$ and $\phi=const$ in (\ref{action-red-compl}).

\subsection{Nearly K\"ahler case}

In the case (\ref{action-nK}) the dimensionally reduced action is
\begin{align} \label{action-red-nK}
S_{nK}[\phi,{\bf a},{\bf e},\rho,\bm{\psi},\bm{\theta},b]=\int -2{\rm Tr}\left( \bm{\psi} d_{\bf a} (\phi {\bf e}^2) + \bm{\theta} (\phi^3 {\bf f} - \phi {\bf e}^2)  - 2\bm{\psi} \bm{\theta} b + \frac{2}{3} \bm{\theta}^3 - \frac{\phi^3}{6} \sqrt{1-\rho^2} \, {\bf e}^3 \right) \\ \nonumber
+ b d(\phi^3).
\end{align} 

As in the complex case, we see that the variation with respect to $\rho$ gives $\rho=0$, and so we can set $\rho$ to zero from the beginning. The other equations are as follows. Varying this action with respect to $b$ we get
\be\label{eqn-nK-1}
d(\phi^3) + 4{\rm Tr}(\bm{\psi} \bm{\theta}) =0.
\ee
Varying with respect to $\bm{\psi}$ we get
\be\label{eqn-nK-2}
d_{\bf a}(\phi {\bf e}^2) = 2\bm{\theta} b.
\ee
Varying with respect to $\bm{\theta}$ gives
\be\label{eqn-nK-3}
\phi {\bf e}^2 - \phi^3 {\bf f} + 2\bm{\psi} b= 2\bm{\theta}^2.
\ee
Varying with respect to ${\bf e}$ gives
\be\label{eqn-nK-4}
d_{\bf a}\bm{\psi} {\bf e} + {\bf e} d_{\bf a}\bm{\psi} + \bm{\theta} {\bf e} + {\bf e} \bm{\theta} + \frac{\phi^2}{2} {\bf e}^2=0.
\ee
Varying with respect to $\phi$ gives
\be\label{eqn-nK-5}
2{\rm Tr}\left( d_{\bf a}\bm{\psi} \phi {\bf e}^2  + \bm{\theta} \phi {\bf e}^2 - 3\phi^3 \bm{\theta} {\bf f} + \frac{\phi^3}{2} {\bf e}^3\right) = 3\phi^3 db.
\ee
Finally, varying with respect to the connection we get
\be\label{eqn-nK-6}
\phi(\bm{\psi} {\bf e}^2 - {\bf e}^2 \bm{\psi}) = d_{\bf a} (\phi^3 \bm{\theta}). 
\ee
There is now no gauge symmetry that can be used to set some of the fields to zero. 

To solve the above system, it seems that the following interpretation of the above equations should be adopted. If one assumes that $\bm{\theta}$ is non-degenerate, then equation (\ref{eqn-nK-1}) determines $\bm{\psi}$ in terms of the derivative of $\phi$, and inverse of $\bm{\theta}$. Then equation (\ref{eqn-nK-2}) determines $b$ in terms of the derivatives of $\phi$ and ${\bf e}$. It is convenient to parametrise ${\bf a}= \bm{ \go} + {\bf t}$, where $\bm{ \go}$ is the connection compatible $d_{\bm{ \go}} {\bf e}=0$ with the frame ${\bf e}$, and ${\bf t}$ is the torsion. Then (\ref{eqn-nK-2})  determines $b$ in terms of $d\phi$ and the torsion, as well as inverse of $\bm{\theta}$. Then equation (\ref{eqn-nK-4}) gives $\bm{\theta}$ in terms of ${\bf e}$ and other data. Equation (\ref{eqn-nK-5}) is the equation on $\phi$. The last equation (\ref{eqn-nK-6}) can be used to determine the torsion ${\bf t}$. Finally, (\ref{eqn-nK-3}) becomes the equation giving ${\bf e}$. 

The simplest solution of this set of equations can be obtained by putting
\be
\bm{\psi}=0, \quad b=0, \quad \bm{\theta} = \alpha {\bf e}, \quad \phi=const.
\ee
In this case ${\bf a}$ is the spin connection compatible with the frame ${\bf e}$, i.e. the torsion ${\bf t}=0$ vanishes. The non-trivial equations are then (\ref{eqn-nK-3}),(\ref{eqn-nK-4}) and (\ref{eqn-nK-5}), and these reduce to 3 algebraic equations for 3 unknowns $\alpha, \phi, \sigma$, where $\sigma$ appears as the curvature ${\bf f} = \sigma {\bf e}^2$. There is a real solution of these equations when $\sigma<0$, which corresponds to positive curvature. A cone over this solution defines a manifold of holonomy $G_2$, and is what the solution of \cite{BS} asymptotes to. We review the \cite{BS} solution in the Appendix. It would be interesting to obtain more general solutions of the above system of equations. 

\section{Discussion}

In this paper we considered the topological theory of 2- and 3-forms $BdC$ in 6 dimensions. We studied the effect of changing this topological Lagrangian by adding potential terms for $B$ and $C$ fields. Three different choices of the potential term were considered, two of them depending on just $B$ or just $C$, and the last one depending on both. In the first two cases we were able to show that the theory remains topological, i.e. that there are no propagating degrees of freedom. Our analysis in the last case was inconclusive, as the algebra of the constraints did not close. Our guess was that there are propagating degrees of freedom in that case. 

From the two topological theories that we described, the more interesting one is (\ref{action-complex}) with the $C$-dependent potential. The critical points of this theory are complex (or para-complex, depending on the type of 3-form one considers) manifolds. This theory can be viewed as a background independent version of Hitchin's theory \cite{Hitchin:2000jd} of 3-forms in a fixed cohomology class. We have shown that its dimensional reduction to 3D gives a version of 3D gravity with non-zero cosmological constant. 

The complex case theory (\ref{action-complex}) can be quantised. The one-loop quantisation of Hitchin's theory was carried out in \cite{Pestun:2005rp}. Most considerations of that paper still apply to (\ref{action-complex}), and so the one-loop partition function of this theory is known. As we have shown in this paper, the dimensional reduction of this theory gives 3D gravity, whose quantisation is also understood, at least for the case of a positive cosmological constant. In particular, the partition function of 3D gravity can be computed via state sum models of \cite{Turaev:1992hq}. Given that this theory is interpreted as the dimensional reduction of the theory (\ref{action-complex}), a very interesting open question is if there is also a state sum model quantisation of the theory (\ref{action-complex}), so that the Turaev-Viro model can be seen arising as the dimensional reduction of some yet to be constructed 6D state sum model. 

Another interesting outcome of this work is the set of equations (\ref{eqn-nK-1})-(\ref{eqn-nK-6}), which is the dimensional reduction of the nearly K\"ahler equations (\ref{feqs-nK}). As we have described, using the cone construction, solutions of this set of equations can be lifted to holonomy $G_2$ structures in 7-dimensions. We have described some simple solutions of this system of equations, corresponding to known \cite{BS} holonomy $G_2$ manifolds. It would be interesting to obtain other solutions. 

The models we considered in this paper were obtained by taking a manifestly topological theory, and deforming it by adding a suitable potential term. Another interesting question is whether it is possible to give a complete list of topological field theories obtainable this way. We leave this to future research. 

\section*{Acknowledgments}

KK was supported by ERC Starting Grant 277570-DIGT, and is grateful to the Max-Planck-Institute for Gravitational Physics (Albert Einstein Institute), Golm (Potsdam) for hospitality while this work has been carried out. YH was supported by a grant from ENS Lyon.

\section*{Appendix}

The goal of this Appendix is to describe the torsionless $G_2$ structure on the $\R^4=\C^2$ bundle over a 3-dimensional manifold $M$ that was first constructed in \cite{BS}. This gives an example of solution of the dimensionally reduced system of equations (\ref{eqn-nK-1})-(\ref{eqn-nK-6}) for the nearly K\"ahler case. 

\subsection*{Nearly K\"ahler structure on $SU(2) \times M$}

We start by describing a solution to the system of equations (\ref{feqs-nK}) on the principal ${\rm SU}(2)$ bundle over a 3-dimensional manifold $M$.

Let us consider the bundle of orthogonal frames on a 3-dimensional Riemannian manifold of constant curvature, i.e. a principale ${\rm SU}(2)$ bundle over $M$. Let ${\bf e}$ be a 1-form on $M$ valued in the space of anti-Hermitian $2\times 2$ matrices. If we assume non-degeneracy of this object, it can be thought of as a frame field for a Riemannian metric on $M$. 

 If $g\in {\rm SU}(2)$ is the fiber coordinate, we can lift ${\bf e}$ to the total space of the ${\rm SU}(2)$ principal bundle by considering $E=g^{-1}  {\bf e} g$. Let $A$ be the lift to the total space of the bundle of an ${\rm SU}(2)$ connection ${\bf a}$ on the base, see (\ref{lift}). We assume that ${\bf a}$ and ${\bf e}$ are compatible and that ${\bf e}$ is the frame field for a constant curvature metric. This corresponds to two equations
\be\label{dE-dW}
d E + W E+EW=0, \qquad F(W) \equiv dW+WW = \sigma EE.
\ee
Here $\sigma>0$ corresponds to a negative curvature metric, and $\sigma<0$ to positive curvature. With this in hand we can introduce
\be\label{B-hat-C}
B= -2{\rm Tr}(EW), \qquad \hat{C} = -2{\rm Tr}(-\sigma E^3-EW^2).
\ee
The related unhatted 3-form is
\be
C= -2\sqrt{-3\sigma}{\rm Tr} \left( \frac{1}{9\sigma} W^3 + WE^2\right).
\ee
An elementary computation using (\ref{dE-dW}) gives
\be\label{Appdx: dB}
dB = -2{\rm Tr}\left( -(WE+EW)W - E(\sigma E^2-W^2) \right) = \hat{C}.
\ee
and
\be
dC= -2\sqrt{-3\sigma} {\rm Tr} \left( \frac{4}{3} E^2 W^2\right)
\ee
Now using
\be
\tau^i \tau^j = - \frac{1}{4} \delta^{ij} \id + \frac{1}{2} \epsilon^{ijk} \tau^k
\ee
we have
\be
{\rm Tr}(\tau^i \tau^j \tau^k\tau^l) = \frac{1}{8} \delta^{ij} \delta^{kl} - \frac{1}{8}\epsilon^{ijs}\epsilon^{skl}.
\ee
This gives
\be
{\rm Tr}(E^2 W^2) = {\rm Tr}(EW){\rm Tr}(EW).
\ee
Thus, $dC$ is a multiple of $BB$, with the proportionality coefficient different from the one in (\ref{feqs-nK}) because of differences in normalisation of $B,C$ here and in the action (\ref{action-nK}). By rescaling $B,C$ simultaneously one obtains a solution of the system (\ref{feqs-nK}). This solution is of the dimensionally reduced form (\ref{C}), (\ref{B-reduction}), and satisfies the dimensionally reduced system of equations (\ref{eqn-nK-1})-(\ref{eqn-nK-6}), as is discussed in the main text. 

\subsection*{The $G_2$ holonomy metric on $ \R \times SU(2) \times M$}

We now review how the above construction relates to the solution \cite{BS}.

Let's consider the associated bundle $\C^2 \times M$ with structure group $SU(2)$. It naturally comes with a Hermitian metric preserved by the structure group and the total space of this bundle then has the structure of a line bundle $\C^2 \times M \to S^3 \times M \simeq SU(2) \times M$. Let $(r, g ) \in \R \times SU(2) \simeq \C^2$ be coordinates along the fiber. 

We can write the following 3-form on $\C^2 \times M$
\be \label{Appdx:def}
\gO = d(r^2)B + r^2 \hat{C}.
\ee
Using (\ref{B-hat-C}) we can rewrite this form as
\be \label{Appdx:def2}
\gO = -2 {\rm Tr}\left(  f^3\;\frac{E^3}{3} + f g^2 \;E \W \left(2rdr W - r^2 WW \right) \right) ,\qquad \text{with} \;f^3= -3\gs r^2 \; \text{and} \; fg^2 =1. 
\ee
This is a stable 3-form in the 7-dimensional space $\C^2\times M$. 

To proceed, we need the expression for the dual 4-form ${}^*\Omega$, as well as for the metric defined by (\ref{Appdx:def2}). This is computed from the following lemma. For a 3-form of the form
\be\label{Appdx:omega-form}
\Omega = - 2{\rm Tr} \left( \frac{1}{3} X^3 + X S\right),
\ee
with anti-Hermitian 1-forms $X$ and $S= \tht w - ww$ 
for some coordinate $x$ and anti-Hermitian 1-forms $w$, the dual 4-form is given by
\be\label{Appdx:dual-form}
{}^*\Omega = - 2{\rm Tr}\left( -\frac{1}{6} S^2 - X^2 S\right).
\ee
The metric $g_\Omega$ defined by (\ref{Appdx:omega-form}) is given by
\be\label{Appdx:metric}
ds_\Omega^2 = - 2{\rm Tr}\left( X\otimes X + w\otimes w\right) + \theta\otimes \theta.
\ee

The 3-form (\ref{Appdx:def2}) is of the form (\ref{Appdx:omega-form}) with 
\be \label{Appdx: co-frame}
X \coloneqq f\;E, \qquad \tht \coloneqq 2 g\; dr, \qquad  w \coloneqq r g\;W.
\ee
Now, the metric $g_\Omega$ has $G_2$ holonomy if $d \gO= 0$, $d {}^*\gO =0$, see e.g. \cite{BS}. The 3-form (\ref{Appdx:def}) is closed as a consequence of $dB=\hat{C}$. In fact, there is a more general solution first obtained in \cite{BS}. Allowing $f,g$ to be arbitrary functions of $r^2$ the 3-form (\ref{Appdx:def2}) is closed if and only if
\be\label{Appdx:system}
fg^2 = const, \qquad (f^3)' + 3\sigma fg^2 =0. 
\ee
We can then always make $fg^2=1$ by rescaling $\Omega$ and redefining $f,g$. Making this choice we get 
\be\label{fg}
f^3 = M- 3\sigma r^2, \qquad fg^2 = 1,
\ee
where $M$ is the integration constant, which can always be chosen to be $M \in \{- 1, 0,1\}$  by rescaling $r$. Our original choice $f^3 = -3\gs r^2$, $fg^2 =1$ is just a particular solution of the system of equations (\ref{Appdx:system}) corresponding to the choice of integration constant $M=0$. Alternatively, the functions as in (\ref{Appdx:def2}) is what the solution for any value of $M$ asymptotes to for large $r^2$. 

Choosing $-3\sigma=1$ we have, for the cone $M=0$ solution $f=r^{2/3}, g=r^{-1/3}$, and the metric (\ref{Appdx:metric}) reads
\be \label{Appdx: cone metric}
ds_\Omega^2 = -2 R^2 {\rm Tr}\left( E^2+ W^2\right) + 9 dR^2, \qquad \text{with}\; R\equiv r^{2/3}.
\ee 
This is a cone metric on $\R\times M^6$. Note that $-2{\rm Tr}(E^2+W^2)$ is the 6D metric constructed from $B,\hat{C}$ following the procedure described in section \ref{subsection: Metric from B,C}. Thus, \eqref{Appdx: cone metric} really is a cone over the metric constructed from $B$ and $\hat{C}$.

We also need to verify the $d {}^*\Omega=0$ equation. Using (\ref{Appdx:dual-form}), the dual 4-form for (\ref{Appdx:def2}) is
\be
{}^*\Omega = - 2{\rm Tr}\left( \frac{2}{3} g^4 r^3 dr W^3 - f^2 g^2 E^2 (2r dr W- r^2 W^2) \right).
\ee
Its exterior derivative is given by
\be
d{}^*\Omega = ( g^4 r^2 \sigma - f^2 g^2 + (f^2 g^2 r^2)') dr^2 (-2) {\rm Tr}(E^2 W^2),
\ee
and the coefficient in brackets vanishes for (\ref{fg}), showing that the form (\ref{Appdx:def2}) with functions $f,g$ given by (\ref{fg}) is closed and co-closed, and thus defines a metric of holonomy $G_2$ on $\R^4 \times M$.


\begin{thebibliography}{99}

\bibitem{Schwarz:1978cn} 
  A.~S.~Schwarz,
  ``The Partition Function of Degenerate Quadratic Functional and Ray-Singer Invariants,''
  Lett.\ Math.\ Phys.\  {\bf 2}, 247 (1978).
  doi:10.1007/BF00406412
  
\bibitem{Schwarz:1979ae} 
  A.~S.~Schwarz,
  ``The Partition Function of a Degenerate Functional,''
  Commun.\ Math.\ Phys.\  {\bf 67}, 1 (1979).
  doi:10.1007/BF01223197
  
\bibitem{Krasnov:2017uam} 
  K.~Krasnov,
  ``Dynamics of 3-Forms in Seven Dimensions,''
  arXiv:1705.01741 [hep-th].

\bibitem{Herfray:2016std} 
  Y.~Herfray, K.~Krasnov and C.~Scarinci,
  ``6D Interpretation of 3D Gravity,''
  Class.\ Quant.\ Grav.\  {\bf 34}, no. 4, 045007 (2017)
  doi:10.1088/1361-6382/aa5727
  [arXiv:1605.07510 [hep-th]].

\bibitem{Hitchin:2000jd} 
  N.~J.~Hitchin,
  ``The Geometry of Three-Forms in Six Dimensions,''
  J.\ Diff.\ Geom.\  {\bf 55}, no. 3, 547 (2000)
  [math/0010054 [math.DG]].

\bibitem{Hitchin:2001rw} 
  N.~J.~Hitchin,
  ``Stable forms and special metrics,''
  math/0107101 [math-dg].
  
\bibitem{Alexandrov:2004cp} 
  B.~Alexandrov, T.~Friedrich and N.~Schoemann,
  ``Almost Hermitian 6-manifolds revisited,''
  J.\ Geom.\ Phys.\  {\bf 53}, 1 (2005)
  doi:10.1016/j.geomphys.2004.04.009
  [math/0403131 [math-dg]].
  
\bibitem{Krasnov:2016wvc} 
  K.~Krasnov,
  ``General Relativity from Three-Forms in Seven Dimensions,''
  arXiv:1611.07849 [hep-th].
  
\bibitem{Hitchin:2004ut} 
  N.~Hitchin,
  ``Generalized Calabi-Yau manifolds,''
  Quart.\ J.\ Math.\  {\bf 54}, 281 (2003)
  doi:10.1093/qjmath/54.3.281
  [math/0209099 [math-dg]].
  
  \bibitem{Grunewald} R.~ Grunewald, "Six-dimensional Riemannian manifolds with a real Killing spinor," Ann.\ Global Anal.\ Geom.\ {\bf 8} 43-59 (1990).
  
\bibitem{Agricola:2014yma} 
  I.~Agricola, S.~G.~Chiossi, T.~Friedrich and J.~H�ll,
  ``Spinorial description of $SU(3)$-and G$_2$-manifolds,''
  J.\ Geom.\ Phys.\  {\bf 98}, 535 (2015)
  doi:10.1016/j.geomphys.2015.08.023
  [arXiv:1411.5663 [math.DG]].
  
  \bibitem{FKMU} Th.~Friedrich, I.~Kath, A.~Moroianu, U.~Semmelmann, "On nearly parallel G2-structures,"  J.\ Geom.\ Phys.\  {\bf 23},259-286 (1997).
  
\bibitem{Pestun:2005rp} 
  V.~Pestun and E.~Witten,
  ``The Hitchin functionals and the topological B-model at one loop,''
  Lett.\ Math.\ Phys.\  {\bf 74}, 21 (2005)
  doi:10.1007/s11005-005-0007-9
  [hep-th/0503083].
  
  \bibitem{BS} R.~ Bryant and S.~ Salamon, "On the construction of some complete
metrics with exceptional holonomy", Duke Math.\ Journ.\ {\bf 53}, 829 (1989).

\bibitem{Turaev:1992hq} 
  V.~G.~Turaev and O.~Y.~Viro,
  ``State sum invariants of 3 manifolds and quantum 6j symbols,''
  Topology {\bf 31}, 865 (1992).
  doi:10.1016/0040-9383(92)90015-A
 
  
  \end{thebibliography}
\end{document}